\documentclass[aps,prapplied,reprint,groupedaddress]{revtex4-1}
\usepackage{amsbsy}
\usepackage{amssymb}
\usepackage{amsmath}
\usepackage{graphicx}
\usepackage{dcolumn}
\usepackage{bm}

\begin{document}


\title{Vapor-cell-based atomic electrometry for detection frequencies below kHz}


\author{Yuan-Yu Jau}
\email[Corresponding author: ]{yjau@sandia.gov}
\author{Carter Tony}
\affiliation{Sandia National Laboratories, Albuquerque, NM 87123, USA}


\date{\today}

\begin{abstract}
Rydberg-assisted atomic electrometry using alkali-metal atoms contained inside a vacuum environment for detecting external electric fields (E-fields) at frequencies $<$ a few kHz has been quite challenging due to the low-frequency E-field screening effect that is caused by the alkali-metal atoms adsorbed on the inner surface of the container. We report a very slow E-field screening phenomenon with a time scale up to $\sim$ second on a rubidium (Rb) vapor cell that is made of monocrystalline sapphire. Using this sapphire-made Rb vapor cell with optically induced, internal bias E-field, we demonstrate vapor-cell-based, low-frequency atomic electrometry that responds to the E-field strength linearly. Limited by the given experimental conditions, this demonstrated atomic electrometer uses an active volume of 11 mm$^3$ and delivers a spectral noise floor around $0.34$ (mV/m)/$\sqrt{\rm Hz}$ and the 3-dB low cut-off frequency around 770 Hz inside the vapor cell. This work investigates a regime of vapor-cell-based atomic electrometry that was seldom studied before, which may enable more applications that utilize atomic E-field sensing technology.
\end{abstract}


\maketitle

\section{Introduction}
The idea of atomic electric-field (E-field) sensors using high-lying Rydberg states of atoms was proposed back in 1990s \cite{Osterwalder1969}. Because of the much larger electric dipole moment associated with a Rydberg state of high principal quantum number $n$, the energy of this high-$n$ Rydberg state can be more easily perturbed by the applied electric field. Hence, sensitive Rydberg-assisted atomic electrometry can in principle be achieved by performing Rydberg spectroscopy to measure the E-field dependent frequency shifts of high-$n$ Rydberg resonances. Similar to other atomic sensing technologies, atomic electrometry can provide not only high sensitivity, but also the best accuracy due to the fact that the E-field driven frequency shifts of Rydberg resonances are only determined by fundamental constants. In addition, the same atoms are indistinguishable at different places. Atomic electrometry can therefore be used as a calibration standard. To date, many E-field measurements using Rydberg states of alkali-metal atoms were demonstrated, and alkali-metal atoms were confined in the vacuum environment using physical containers, such as vacuum chambers and vapor cells. Despite the reported E-field sensitivity values that show great potential of Rydberg-assisted atomic electrometry, due to the low-frequency E-field screening effect \cite{Mohapatra2007,Mohapatra2008,Viteau2011,Cox2018} from the alkali-metal-atom container, E-field signals that originate outside the vacuum environment could be detected only if that electric field is oscillating at high frequencies (from sub-MHz to THz) \cite{Sedlacek2012,Sedlacek2013,Holloway2014,Fan2015,Miller2016,Holloway2017,Meyer2018,Wade2018,Paradis2019,Holloway2019,Meyer2020}. Hence, detecting electric fields with frequencies $<$ a few kHz had only been achieved with the E-field sources that is also inside the vacuum environment \cite{Kubler2010,Abel2011,Carter2012,Hankin2014,Facon2016}. From a practical point of view, atomic electrometry can fill up more application space if we can extend the detection spectrum of the electric field originated outside the vacuum environment to low-frequency and static (DC) regions.

In this paper, we show that a rubidium (Rb) vapor cell made of monocrystalline sapphire can have a very slow time scale of E-field screening up to $\sim$ second. Using a sapphire-made vapor cell with optically induced, internal bias E-field, we are able to demonstrate low-frequency atomic electrometry that is sensitive enough to see weak ambient electric-field signals from the AC power and other electronic noise and to remotely detect moving charged objects. To our best knowledge, vapor-cell-based atomic electrometry at this low-frequency range with our demonstrated sensitivity has not been reported. Our work may enable new atomic E-field sensing applications, such as calibrating magnitudes of DC and low-frequency electric fields, non-invasive diagnostics of electronics in extremely low current mode that has only voltage signatures, communications in ELF and SLF bands ($\lesssim$ kHz), proximity sensing, detection of remote activities, geoscience studies, and bioscience studies.

\section{E-field screening effect on a vapor cell}
\begin{figure}[t]
\includegraphics[trim=70 550 0 40,angle=0,scale=0.78]{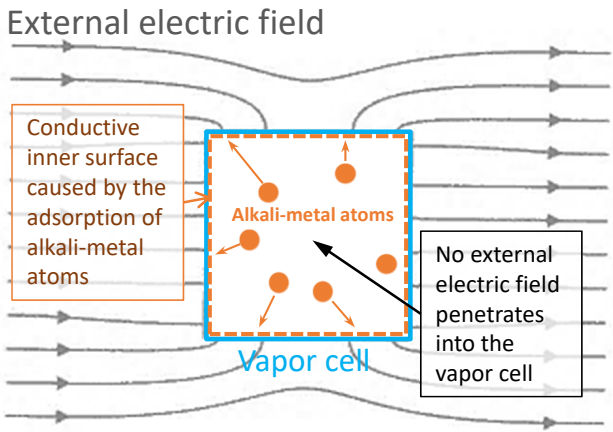}
\caption{\label{VCES}Adsorption of alkali-metal atoms causes inner surface of a vapor cell to be slightly conductive. When low-frequency external electric field is applied, free charges on the inner surface redistribute to maintain equal potential, and no electric field leaks into the vapor cell.}
\end{figure}

Glass materials for making vapor cells are usually good electric insulators. Because of the adsorption of alkali-metal atoms, the inner surface of a vapor cell can have non-zero conductivity. As illustrated in Fig.~\ref{VCES}, when the slowly-varying external electric field is applied, the surface free charges redistribute to maintain equal potential on the conductive surface and null the electric field that is originated externally. The E-field screening rate is determined by the speed of redistributing surface free charges. Furthermore, when a vapor-cell is exposed to laser beams, more surface free charges can be generated through photoelectric effect to enhance the screening rate. To have a more quantitative understanding, we can consider an analytically solvable case: a spherical vapor cell with a radius $r$ and negligible glass thickness, and we find $E_i(t)=E_e\exp(-t/\epsilon R_\Box r)$, where $E_i(t)$ is the time-dependent, internal E-field amplitude caused by the externally applied, uniform electric field, which suddenly turns on at $t=0$ and has an amplitude $E_e$, $\epsilon$ is the effective electric permittivity of the space, and $R_\Box$ is the sheet resistance on the inner surface. This time response leads to a high-pass filtering behavior in the frequency domain. For example, in some common situation, $\epsilon\approx10^{-11}$ F/m, $R_\Box<10^8$ $\Omega/\text{sq}$, and $r\approx0.01$ m, we find the E-field screening time constant to be $\lesssim10^{-5}$ s, and therefore the 3-dB low cut-off frequency, $f_{\rm 3dB}$, is $\gtrsim10^4$ Hz. In addition, we find that to shield 1 V/m electric field at the frequency below $f_{\rm 3dB}$, it only requires free charges on the order of $10^{-14}$ C. From first-order approximation, if the vapor-cell dimension is roughly isotropic in size, the time scale of E-field screening is on the order of $\epsilon R_\Box V^{1/3}$, where $V$ is the volume of the vapor cell. On the other hand, the time response of the E-field screening effect can always be precisely calculated with numerical models depending on the vapor-cell geometry.

In order to further reduce the E-field screening rate for low-frequency sensing, we can increase $\epsilon$, $R_\Box$, and $V$. Usually there is not much room for tuning the electric permittivity, and increasing $V$ leads to a larger physical dimension of the sensor. Hence it is more preferable to increase the inner surface sheet resistance $R_\Box$. According to the results from the earlier studies \cite{Sakai1977,Bouchiat1999}, Al$_2$O$_3$ materials not only show good resistance to the corrosion by alkali-metal atoms but also demonstrate higher surface sheet resistance compared to the SiO$_2$-based glass materials when exposing to alkali vapor. To verify the results from that prior work, we experimentally characterized Rb vapor cells made of fused silica and monocrystalline sapphire. From our best experimental results, with the same internal vapor density, the sapphire-made Rb cell demonstrates $R_\Box$ on the inner surface that is orders of magnitude higher than what we can obtain from fused-silica Rb vapor cells. With a sapphire-made Rb cell, we find $R_\Box>10^{12}\Omega/\text{sq}$ with no exposure to laser beams and $R_\Box>10^{9}\Omega/\text{sq}$ with presence of high-power laser beams.

\begin{figure*}[t]
\includegraphics*[trim=0 475 0 0,angle=0,scale=0.78]{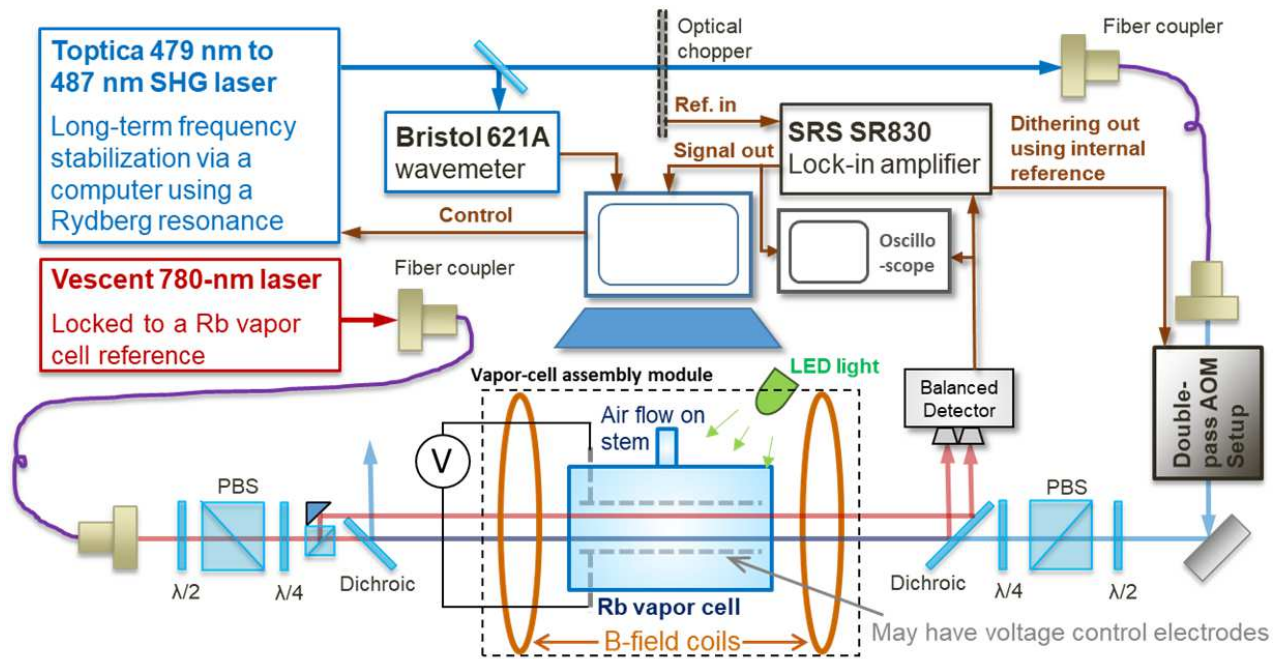}%
\caption{\label{Exp}Overview of the experimental implementation for measurements of E-field dependent Rydberg resonances, characterizations of E-field screening effect on vapor cells, and low-frequency atomic electrometry.}
\end{figure*}
\section{Experimental implementation}
For the work desribed in this paper, there are three major experimental tasks: 1. Verify the modeling results of E-field dependent Rydberg resonances by comparing with experimental measurements; 2. Use E-field dependent Rydberg resonances to characterize E-field screening effects of Rb vapor cells; 3. Perform low-frequency atomic electrometery using a vapor cell that demonstrates the highest $R_\Box$ with internal bias E-field generated by optical means. The overall implementation for these experiments is illustrated in Fig.~\ref{Exp}.

We probed Rydberg states of Rb atoms inside vapor cells using a scheme of stimulated Raman transition that results in a phenomenon of electromagnetically induced transparency (EIT), which has been widely used for optically detecting Rydberg states of alkali-metal atoms \cite{Mohapatra2007,Zhao2009}. There were two laser sources at 780 nm and $\sim480$ nm for EIT interrogation of Rb Rydberg states. The 780-nm laser was locked to one of the $5S_{1/2}$ to $5P_{3/2}$ transitions. With the assistance of a wavemeter, the 480-nm laser was tuned to a selected $nS$ or $nD$ level from the $5P_{3/2}$ level based on the calculated optical transition frequencies. Two polarization maintaining (PM) fibers were used to deliver the laser light to the vapor-cell apparatus. Two laser beams from the output ports of the optical fibers were aligned to point toward each other collinearly. The polarization optics were used to optimize the signal. For each experiment, a vapor cell was mounted by a 3D-printed fixture and placed in the beam path. The magnetic-field (B-field) coils were wound on 3D-printed coil frames. The vapor cells used in the experiments were made of either fused silica or sapphire and filled with natural abundance Rb metal. For studying the E-field dependent frequency shifts of Rydberg resonances, we used a fused silica cell with internal, parallel metal plates connected to a voltage source through electric feedthroughs to produce electric field with a desired strength inside the vapor cell. All the vapor cells were pre-treated by heating up the cell body and keeping the cell stem (cold finger) cooled to drive the Rb metal into the stem and to clean up the inner surface of the cell body, which is a necessary scheme for minimizing the E-field screening effect regardless the vapor-cell materials. We controlled the cell-body temperature by wrapping the vapor cell with heating wires driven by electric current. The cell stem was blown by cold air with adjustable flow rate. We used miniature thermocouples to measure and control the cell-body and the cell-stem temperatures. Once the Rb metal was collected into the stem, the vapor density can be controlled by the stem temperature, as long as it is below the cell-body temperature. The atom number density of Rb vapor was determined by measuring the absorption spectrum at 780 nm. Depending on the requirement of the specific experiment, we can choose to keep or dismiss the heating wires on the vapor cell.

For probing Rydberg signals, we detected the transmission of the 780-nm beam through a vapor cell and modulated the 480-nm laser either in intensity or in optical frequency. An optical chopper was used for intensity modulation at a few kHz rate, and a double-pass AOM (acoustic-optical modulator) setup was used for dithering the optical frequency at a few tens of kHz rate. The dithering range was usually set to match with the Rydberg resonance linewidth. We used a lock-in amplifier to demodulate and reveal the Rydberg signals from the 780-nm detection. The original resonance lineshape can be observed when modulating the 480-nm laser intensity, and when modulating the optical frequency, a dispersive-like resonance signal can be observed. A balanced detector that received the transmissions of the main 780-nm beam (overlapped with the 480-nm beam) and a reference 780-nm beam was used for minimizing the detection noise, which was mainly caused by the intrinsic intensity fluctuation of the 780-nm laser and additional intensity noise converted from the frequency/phase noise of the 780-nm laser via the Rb vapor cell \cite{Camparo1999} for an experiment. We used oscilloscopes for real-time signal monitoring and quick data acquisition. A computer was used for detailed data acquisition, 480-nm laser control, and 480-nm laser stabilization with a slow feedback loop using a Rydberg resonance when needed.

\section{E-field dependent Rydberg Resonances}
Understanding how the applied electric field affects the frequencies and the amplitudes of Rydberg resonances in the presence of a bias B-field is necessary for conducting low-frequency atomic electrometry. We carried out both modeling work and experimental measurements to study E-field dependent Rydberg resonances. We calculate the energy shifts of the Rydberg states and the associated sublevels by solving the eigenvalues of the total Hamiltonian $H=H_0+H_E+H_B$, where $H_0$ is the unperturbed Hamiltonian, and $H_E$ and $H_B$ account for the interactions with electric field and magnetic field via electric and magnetic dipole moments. We used experimentally determined quantum defects \cite{Li2003,Afrousheh2006,Han2006,Lee2016} to calculate the energies of unperturbed, Rb Rydberg states labeled by $(n,l,j)$, where $n$ is the principal quantum number, $l$ is the orbital quantum number, and $j$ is the total angular momentum quantum number of a fine-structure multiplet due to the spin-orbit coupling. Hence, $H_0$ can be constructed. With the given state energies, the valence electron radial wavefunction of each $(n,l,j)$ state then can be calculated using Coulomb approximation or model potentials \cite{Happer2010}. By combining the radial matrix elements derived from the wavefunction calculations and the spherical harmonics for orbital angular momentum states, we can figure out the matrix elements of the electric-dipole-moment operator of the associated atomic states. We can therefore construct $H_E$. The Hamiltonian $H_B$ for B-field interaction can be easily generated by the angular-momentum operators with the associated gyro-magnetic ratios. In addition to the frequency shifts of Rydberg resonances calculated by $H$, we simulate the relative signal strengths of the Rydberg resonances on EIT spectroscopy using full-level density-matrix modeling \cite{Happer2010}, which allows us to take account of the effects from the actual experimental parameters, such as the polarization of the laser beams, orientation of the applied E-field, the direction of the bias B-field, etc. Similar numerical calculations for Rydberg spectroscopy with single excitation laser were performed in the prior work \cite{Hankin2014,Jau2016} for cesium atoms.

\begin{figure}[t]
\includegraphics[trim=0 200 0 0,angle=0,scale=0.7]{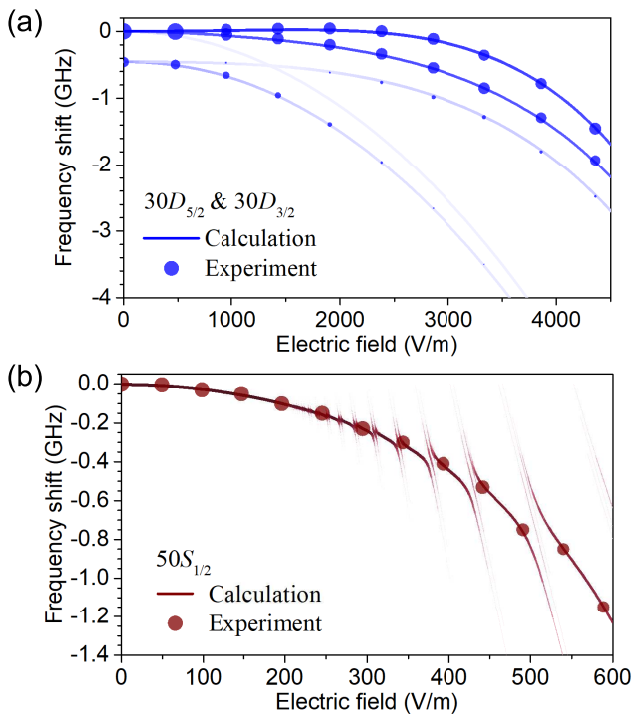}
\caption{\label{Eshifts}E-field dependent frequency shifts of Rb Rydberg resonances. Curves represent the modeling results, and the resonance strengths are indicated by the darkness of the color. Filled circles are the experimental data points, and the relative signal strengths are indicated by the size of the circle.}
\end{figure}
In the experiments of EIT Rydberg spectroscopy, we used a cylindrical Rb vapor cell made of fused silica with parallel metal plates inside. The cell was made by Precision Glassblowing. It was about 50 mm long and has 25 mm OD (outer diameter). The internal metal plates were $\approx2$ mm apart, roughly 22 mm wide and 44 mm long, and were made of stainless steel. The 780-nm laser beam was 1 mm in diameter $(1/e^2)$ with 20 $\mu$W optical power. The 780-nm laser was locked to the $F=3$ to $F'=4$, $^{85}$Rb $D2$ transition. The 480-nm laser beam was 1 mm in diameter $(1/e^2)$ and chopped at 3.9 kHz, 50\% duty cycle, with 50 mW averaged optical power. The laser beams were transmitted through the region between the metal plates inside the cell, and the E-field inside this region was generated by connecting the metal plates to a stable voltage source via electric feedthroughs on the cell. Hence, the E-field is perpendicular to the laser beams. A bias B-field at 2 G (0.2 mT) along the beam direction was added by the B-field coil sets . Figure~\ref{Eshifts} illustrates some experimental data compared with the modeling results. For Fig.~\ref{Eshifts}(a), the 480-nm laser was tuned to reach $D$ orbital with principal quantum number $n=30$ and slowly swept its optical frequency for a range of a few GHz. Both laser beams were linearly polarized, and the polarization was aligned to the E-field. The resonance signals were revealed at the output of the lock-in amplifier, and the resonance frequency was determined by the resonance peak position. There are two resonances at zero E-field, labeled by $30D_{5/2}$ at 0 GHz and $30D_{3/2}$ at -0.45 GHz frequency offsets. By increasing the E-field, resonances split into more resonances, three from $30D_{5/2}$ and two from $30D_{3/2}$, with different signal strengths. We determined the signal strength by measuring the area under each resonance owing to the fact that the resonance width increased when the E-field was getting much stronger. The linewidth broadening was mainly caused by the residual E-field inhomogeneity that is proportional to the overall E-field strength. We see that the modeling results match pretty well with the experimental data. The resonance of the lowest frequency from $30D_{5/2}$ was not measured and is hardly seen in the plot due to its much weaker strength and the color scheme for plotting the curves. In Fig.~\ref{Eshifts}(b), we plot the calculated frequency-shift curves and the experimentally measured data points for $50S_{1/2}$ state. We find many anti-crossing features on the curve for the electric-field strength above 200 V/m, which were also observed from the EIT spectroscopy. This is due to the energy levels extended from the Rydberg states of $n=47$ and $l=3$ to $l=46$ that interfere with the $50S_{1/2}$ state. In this measurement, both 780-nm and 480-nm laser beams were circularly polarized.

\section{Characterization of Sapphire-made R\lowercase{b} vapor cell}
\begin{figure}[b]
\includegraphics*[trim=40 35 345 0,angle=-90,scale=0.75]{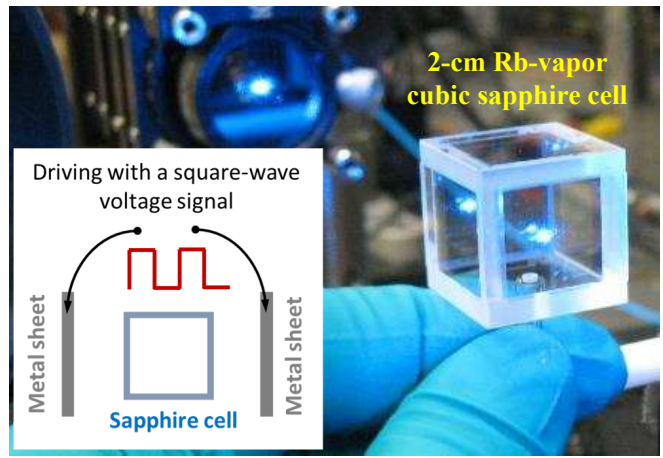}%
\caption{\label{SapphireCell} Picture of the sapphire-made Rb vapor cell. The inset illustrates the method of producing the external electric field switching between two polarities for characterizing the E-field screening effect on the cell.}
\end{figure}
\begin{figure}[t]
\includegraphics*[trim=0 15 0 0,angle=0,scale=0.7]{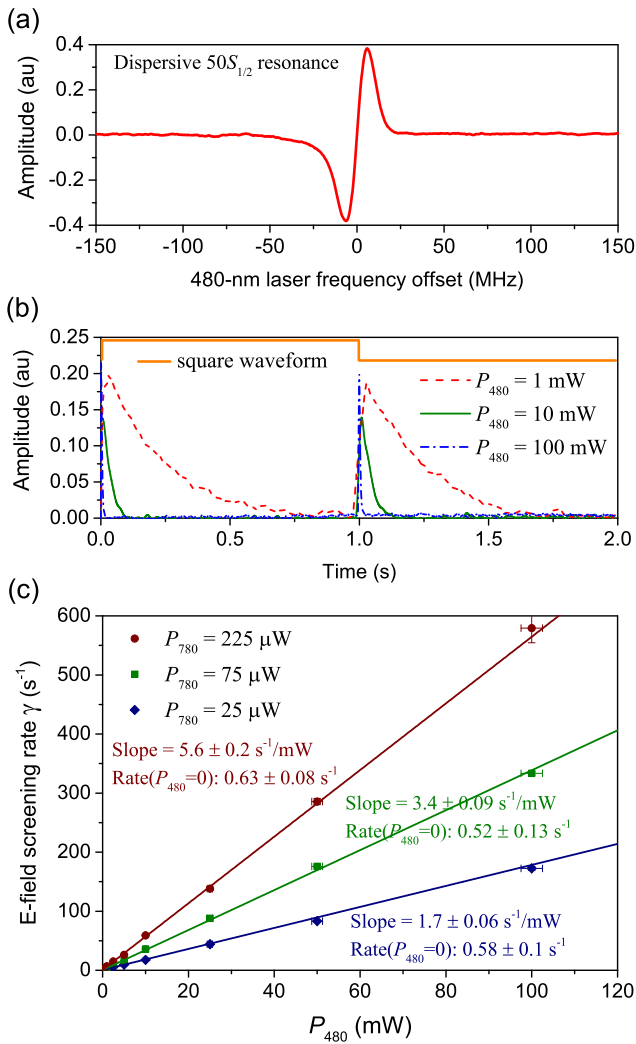}%
\caption{\label{ScreenningRates}(a) A typical dispersive-like Rydberg resonance signal of $50S_{1/2}$. (b) An example of time transients of Rydberg resonance shifts using different $P_{480}$ values. (c) E-field screening rates of a sapphire cell with stem temperature at 23$^\circ$C and body temperature at 100 $^\circ$C versus different $P_{780}$ and $P_{480}$ values.}
\end{figure}
To characterize the E-field screening effect of a Rb vapor cell made of sapphire, we applied sudden changes to the electric field externally to the cell and monitor the frequency shift of the $50S_{1/2}$ Rydberg resonance as a function of time. The sapphire cell used in the experimental characterization was made by Japan Cell Inc. with two C-Cut windows for laser beam access. The cell was in a cubic geometry with an outer dimension of 2 cm along each side. The thickness of the cell windows is 3 mm. Hence, the internal volume is $1.4^3$ cm$^3$. Figure~\ref{SapphireCell} shows a picture of the cubic sapphire cell with laser beams passing through and illustrates the means of producing external electric field switching in time by driving the two metal sheets (copper foils) on each side of the cell with a square-waveform voltage signal. In the experiments, the vapor density was controlled by the cell-stem temperature, which was fixed at 23 $^\circ$C. The cell body was heated by using high-resistance heating filaments, which used very small portion of the cell outer surface area. The entire cell body was enclosed by polyimide sheets as the thermal insulator, and only the stem was left outside. The temperature values were measured by miniature thermocouples. The cell-body temperature was set at 23 $^\circ$C, 50 $^\circ$C, and 100 $^\circ$C. The atom number density of $^{85}$Rb was measured by a 780-nm laser probe, and the results were consistent with the calculations using the vapor-pressure formula with the stem temperature.

To probe the Rydberg resonance, the 780-nm laser beam was 2 mm in diameter $(1/e^2)$, and the 480-nm laser beam was 1 mm in diameter $(1/e^2)$. We used three different 780-nm laser power values, $P_{780}=$ 25, 75, and 225 $\mu$W and several 480-nm laser power values, $P_{480}$ from 0.1 to 100 mW for the characterization work. We used a double-pass AOM to dither the optical frequency of the 480-nm laser beam and used a lock-in amplifier for demodulation to generate a dispersive-like resonance signal as shown in Fig.~\ref{ScreenningRates}(a). The zero-crossing of the signal defines the resonance frequency. At each time of switching the external electric field, a sudden jump of the Rydberg resonance frequency occurs and the following transient indicates the redistribution of the free charges on the inner surface of the vapor cell over a certain time scale, which ends up with equal electric potential on the inner surface. The external electric field is then shielded. Figure~\ref{ScreenningRates}(b) illustrates the $50S_{1/2}$ resonance frequency evolving as a function of time by switching the external electric field with three different $P_{480}$ values, and $P_{780}$ was fixed at 25 $\mu$W. By setting the 480-nm laser frequency at the resonance frequency of zero electric field and taking advantage of the linear response at zero-crossing of the dispersive resoance, the frequency shift of the resonance is converted to the amplitude response. At both positive and the negative electric-field transitions, we see the resonance shifts to the same sign that is because of the quadratic E-field dependence (i.e. the frequency shift $\nu_s\propto E_i^2$) near zero electric field ($E_i\approx0$). We averaged the transient data from both the positive and the negative periods of the square wave to reduce systematic effects and fit the data with an exponential decay function that is proportional to $\exp(2\gamma t)$ to obtain the E-field screening rate $\gamma$. The factor of 2 in front of $\gamma$ is due to the quadratic dependence of the E-field. For different $P_{780}$ and $P_{480}$ values, we used different lock-in gains, lock-in time constants, and data averaging times to maintain similar signal size and signal-to-noise ratio (SNR).

\begin{table}[b]
   \centering 
   \begin{tabular}{l|c|c|c|c} 
   \hline
   Values \& $P_{780}$ & 23 $^\circ$C & 50 $^\circ$C & 100 $^\circ$C & unit\\
   \hline
    $\gamma/P_{480}$, 25$\mu$W & $3.2\pm0.08$ & $2.0\pm0.03$ & $1.7\pm0.06$ & s$^{-1}$/mW\\
    $\gamma/P_{480}$, 75$\mu$W & $4.3\pm0.1$ & $3.7\pm0.06$ & $3.4\pm0.09$ & s$^{-1}$/mW\\
    $\gamma/P_{480}$, 225$\mu$W & $5.7\pm0.2$ & $5.8\pm0.2$ & $5.6\pm0.2$ & s$^{-1}$/mW\\
    \hline
    $\gamma_0$, 25$\mu$W & $1.2\pm0.05$ & $0.75\pm0.05$ & $0.58\pm0.1$ & s$^{-1}$\\
    $\gamma_0$, 75$\mu$W & $1.2\pm0.06$ & $0.79\pm0.05$ & $0.52\pm0.13$ & s$^{-1}$\\
    $\gamma_0$, 225$\mu$W & $1.2\pm0.08$ & $0.81\pm0.08$ & $0.63\pm0.08$ & s$^{-1}$\\
   \hline
   \hline
   $R_{\Box,0}$ & $2.3\pm0.2$ & $3.5\pm0.4$ & $4.7\pm1$ & $10^{12}\Omega/\text{sq}$\\
   \hline
   \end{tabular}
   \caption{\label{CS}Summary of Rb sapphire cell characterization with three different cell-body temperatures. Here $\gamma_0$ and $R_{\Box,0}$ represent the E-field screening rate and $R_{\Box}$ with $P_{\rm 480}=0$.}
\end{table}
We summarize the measured E-field screening rates with different experimental conditions of a Rb sapphire cell in Table~\ref{CS}. Figure~\ref{ScreenningRates}(c) plots an example data set for cell-body temperature at 100 $^\circ$C. We clearly see that the E-field screening speed is proportional to $P_{480}$ when other parameters are fixed. This is likely ascribed to the free electrons excited by the 480-nm laser on the inner surface of the cell. In addition, with the presence of the 480-nm laser, higher $P_{780}$ also leads to a higher screening rate, but it is not linear. The increase of the screening rate gets weaker at much higher $P_{780}$. However, we do not see significant effect from $P_{780}$ with $P_{480}=0$. Hence, 780-nm photons may not have enough energy to directly produce free electrons on the surface. The reason that 780-nm laser can affect the screening rate may be associated with the excited Rb atoms produced by both 780-nm and 480-nm photons. But the detailed mechanism of generating surface free charges from the excited Rb atoms could be complicated, and it is beyond the scope of this work. The 4th to 6th rows in Table~\ref{CS} list the screening rates, $\gamma_0$, by extrapolating data points to $P_{480}=0$. For the same cell-body temperature, the measured $\gamma_0$ is basically the same regardless of the 780-nm laser power. From the finite-element modeling (FEM) for this sapphire cell, we find $R_\Box=2.72\times10^{12}/\gamma$ $\Omega/\text{sq}$. We use the averaged values from 4th to 6th rows for each temperature to calculate the values of the intrinsic, inner surface sheet resistance $R_{\Box,0}$ (without optical illumination to the cell), which are listed in the last row of Table~\ref{CS}. We find that hotter surface temperature leads to higher $R_{\Box,0}$, which may be explained by the temperature-dependent adsorption of Rb atoms on the surface.

When fitting the $1/R_{\Box,0}$ values of the three cell-body temperatures using a function $\propto\exp(\mathcal{E}/kT)$, where $k$ is the Boltzmann constant and $T$ is the absolute temperature of the cell body, we find $\mathcal{E}\approx0.11$ eV, which is much smaller than the reported adsorption energy, 0.7 eV, of Rb atoms on sapphire \cite{Petrov2017}. On the other hand, when using the bulk resistivity of Rb, we find that we need a rubidium film with thickness much less than one atomic monolayer to achieve the measured $R_{\Box,0}$. This indicates the surface Rb atoms are extremely dilute, which was also pointed out by the prior work \cite{Bouchiat1999}. Hence, the mechanism of the inner-surface electrical conduction of the sapphire cell is quite different from the bulk Rb metal. The electron energy band structure can be complicated. Therefore, the surface conductivity is not only proportional to the Rb atom number on the surface, which is determined by the adsorption energy, but also determined by an excitation probability from a lower energy band or state. Hence, we have $1/R_{\Box,0}\propto\exp(\mathcal{E}/kT)$, and $\mathcal{E}=\mathcal{E}_a-\mathcal{E}_g$, where $\mathcal{E}_a$ is the adsorption energy and $\mathcal{E}_g$ is the gap energy for excitation, and we find $\mathcal{E}_g\approx0.59$ eV.

We used similar characterization method for Rb cells made of fused silica that were about the same size as the sapphire cell. For the same stem and the cell-body temperatures, we found that the best results of the fused-silica cells to be $R_{\Box,0}=(1.8\pm0.2)\times10^9$ $\Omega/\text{sq}$ and $\gamma/P_{480}=(68\pm0.6)$ s$^{-1}$/mW compared to a sapphire cell that has the best values of $R_{\Box,0}=(4.7\pm1)\times10^{12}$ $\Omega/\text{sq}$ and $\gamma/P_{480}=(1.7\pm0.06)$ s$^{-1}$/mW. The intrinsic E-field screening time scale of a fused-silica cell is on the order of $10^{-3}$ s compared to $\sim$ second from a sapphire cell. In our preliminary study, we also found that the E-field screening rate was positively affected by the Rb vapor density and also the 780-nm intensity under our experimental conditions.

\section{Low-frequency atomic electrometry}
\begin{figure*}[t]
\includegraphics*[trim=10 370 0 0,angle=0,scale=0.82]{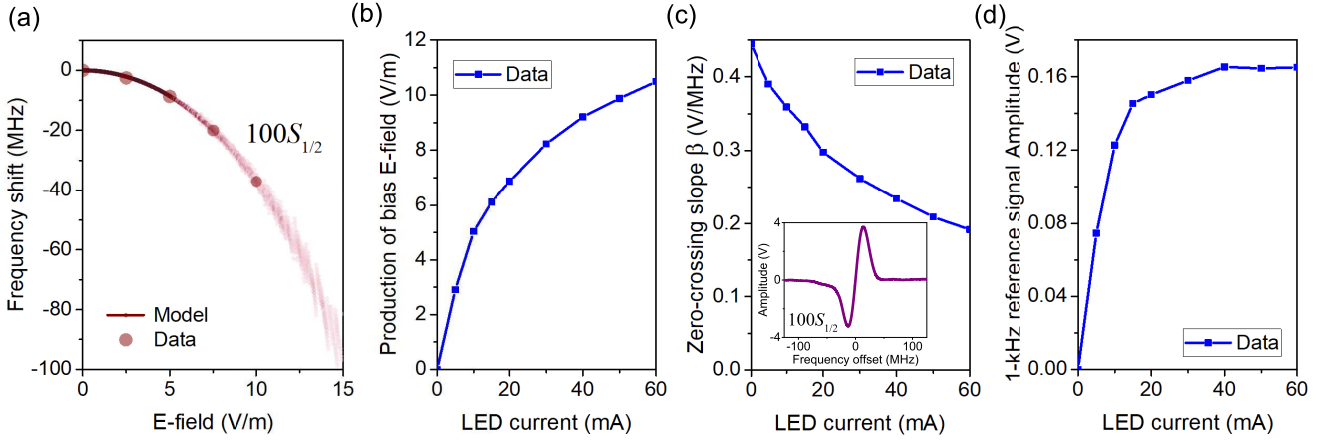}%
\caption{\label{LEDData}(a) Modeling and experimental results of E-field dependent $100S_{1/2}$ resonance shift. (b) Amplitude of the optically induced bias electric field inside the sapphire cell versus $I_{\rm LED}$. (c) The slope at the zero crossing of the dispersive $100S_{1/2}$ resonance versus $I_{\rm LED}$. The inset shows the corresponding dispersive resonance at zero bias E-field. (d)The measured signal amplitude of the E-field reference signal at 1 kHz versus $I_{\rm LED}$.}
\end{figure*}
Taking advantage of the much lower E-field screening rate on the sapphire cell, we demonstrated  vapor-cell-based, low-frequency atomic electrometry. In this work, we chose $100S_{1/2}$ Rydberg state for simple resonance structure and better E-field sensitivity. Figure~\ref{LEDData}(a) shows the E-field dependent shift of the $100S_{1/2}$ resonance from the experimental data using a vapor cell with electrodes and the modeling result. The smearing feature on the shift curve at higher electric-field strength is caused by the energy levels extended from the Rydberg states of $n=97$ and $l=3$ to $l=96$, which leads to effective broadening of the resonance linewidth. The E-field dependent frequency shift $\nu_s$ can be written as $\nu_s=\alpha E_i^2$, where $\alpha\propto n^7$ is a quadratic shift coefficient. In frequency domain, we have $\tilde{E}_i(f)=\tilde{\eta}(f)\tilde{E}_e(f)$, the electric field amplitude inside the cell. Here, $|\tilde{\eta}(f)|=f/\sqrt{f^2+f^2_{\rm 3dB}}$ is the absolute high-pass E-field screening factor, and $\tilde{E}_e(f)$ is the external electric field amplitude.

Different from the experiment of characterizing the E-field screening effect on the sapphire cell, we removed the heating wires and the thermal insulator from the cell and only used an airflow blowing on the cell stem to maintain the stem temperature to be lower than the cell-body temperature, and the whole experimental setup was inside an enclosure. Because of the heat generated by the nearby B-field coils, the lowest researchable cell-stem temperature was hotter than without the enclosure with the same airflow rate, and the measured $^{85}$Rb atom number density was a few times higher than the case in the characterization experiment. The two laser beams were both 1 mm in diameter $(1/e^2)$ with $P_{780}=200$ $\mu$W and $P_{480}$ up to 120 mW. Both laser beams were circularly polarized. The longitudinal bias B-field was set at 6 G. Although the $\alpha$ of $100S_{1/2}$ is about $2^7$ times higher than the $\alpha$ of $50S_{1/2}$, $\nu_s$ is still proportional to $E_i^2$ near zero electric-field background. Assuming the minimum detectable frequency shift is $\delta\nu_s\propto\text{SNR}^{-1}$, we find the minimum detectable electric-field amplitude to be $\propto\sqrt{\delta\nu_s}\propto\text{SNR}^{-1/2}$. This quadratic dependence makes it harder to detect smaller electric field by increasing the SNR. Besides, it makes the low-frequency E-field signal more vulnerable to the E-field screening effect. Since the high-pass shielding factor $\tilde{\eta}\propto f$ when the external electric-field frequency $f$ is much below $f_{\rm 3dB}$, $\nu_s$ is then proportional to $f^2$ for $f\ll f_{\rm 3dB}$. Hence the lower frequency detection is even more difficult. To remove these quadratic disadvantages, we implemented a green LED (light emitting diode) as shown in Fig.~\ref{Exp} that shines focused light onto the cell at a specific spot. With this, we were able to generate an optically induced charged patch at a desired spot, which produced bias E-field inside the cell. Similar phenomenon was reported in Ref.\cite{Hankin2014}. One may also wonder how the charged patches generated by the 480-nm laser light affected the E-field inside the cell. We did not observe significant resonance shift by increasing the 480-nm laser power. We believe that the collimated 480-nm laser beam passing through the two cell windows produced similar strength of the charged patches on each side of the vapor cell and minimize the associated E-field in between the two cell windows. In addition, the LED light did generate surface free charges as well, but it was negligible at normal operating 480-nm laser power and became noticeable only when $P_{480}<1$ mW. In Fig.~\ref{LEDData}(b), we plot the strength of the bias E-field, $E_b$, versus the driving current $I_{\rm LED}$ to the LED, where $E_b$ was determined via the frequency shift of the $100S_{1/2}$ resonance using the frequency shift curve shown in Fig.~\ref{LEDData}(a). Now the internal electric field is $\tilde{E}_i=E_b+\tilde{\eta}\tilde{E}_e$, and the frequency shift is then $\tilde{\nu}_s=\alpha(E_b^2+2\tilde{\eta} E_b\tilde{E}_e+\tilde{\eta}^2\tilde{E}_e^2)$. For small external electric field, we have a linear response of $\tilde{E}_e$ to the frequency shift with a factor $\kappa=2\alpha\tilde{\eta} E_b$. Here $\kappa$ is basically the slope of the curve at E-field strength = $E_b$ in Fig.~\ref{LEDData}(a) when $\tilde{\eta}=1$.

To perform the atomic electrometry, we locked the 480-nm laser to the zero crossing of the dispersive-like $100S_{1/2}$ resonance with a loop time constant on the order of seconds. Hence, we were able to detect electric-field signals with frequencies down to $\sim$ Hz. Our 480-nm laser system was stable enough over the period of the loop time constant. The changes of the electric-field strength was converted to the changes of the resonance frequency that was then converted to an amplitude signal via the slope $\beta$ at zero crossing of the dispersive resonance, the same mechanism we used for characterizing the E-field screening effect. The electrometry signal amplitude is then $\tilde{V}_{\rm sig}=\beta\kappa \tilde{E}_e$, and the E-field sensitivity is $|\tilde{V}_{\rm noise}/\beta\kappa|$, where $\tilde{V}_{\rm noise}$ is the spectral noise. Figure~\ref{LEDData}(c) plots the measured $\beta$ at the zero crossing of the dispersive $100S_{1/2}$ resonance signal. We find that $\beta$ decreases with increased LED current $I_{\rm LED}$. This is caused by the linewidth broadening due to the smearing of the $100S_{1/2}$ shift curve illustrated in Fig.~\ref{LEDData}(a) and mainly due to the spatial gradient of $E_b$ across the probed Rb atoms inside the cell. Therefore, even $\kappa\propto E_b$, but $\beta$ is eventually proportional to $\kappa^{-1}$, and we cannot improve the E-field sensitivity by further increasing $E_b$. On the other hand, this also indicates that increasing $\kappa$ by going to a Rydberg state of higher $n$ will not help the sensitivity either. Figure~\ref{LEDData}(d) plots the amplitude of the detected electric-field signal generated by the parallel copper sheets (made by copper tapes) that were outside the cell and driven by a sinusoidal voltage source at 1 kHz as a reference signal. We see that the signal amplitude increases rapidly when adding the bias E-field, but the increasing starts getting saturated after $I_{\rm LED}\gtrsim20$ mA due to the resonance linewidth is eventually proportional to $E_b$.

\begin{figure}[t]
\includegraphics*[trim=30 10 0 50,angle=0,scale=0.35]{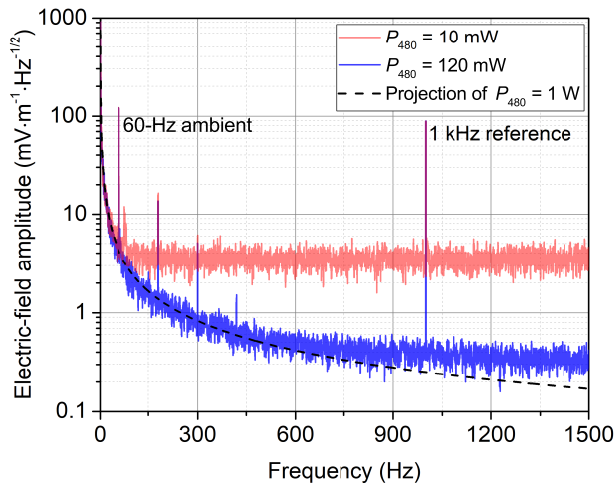}%
\caption{\label{Sensitivity}Measured electric-field spectrum of atomic electrometry with $P_{780}$ at 200 $\mu$W and $P_{480}$ at 10 mW and 120 mW. Higher 480-nm laser power improves the E-field sensitivity at higher frequency, but the sensitivity for frequency $\ll$ $f_{\rm 3dB}$ remains the same. The dashed curve is the projection of using $P_{480}=1$ W.}
\end{figure}
\begin{figure*}[t]
\includegraphics*[trim=30 480 0 0,angle=0,scale=0.9]{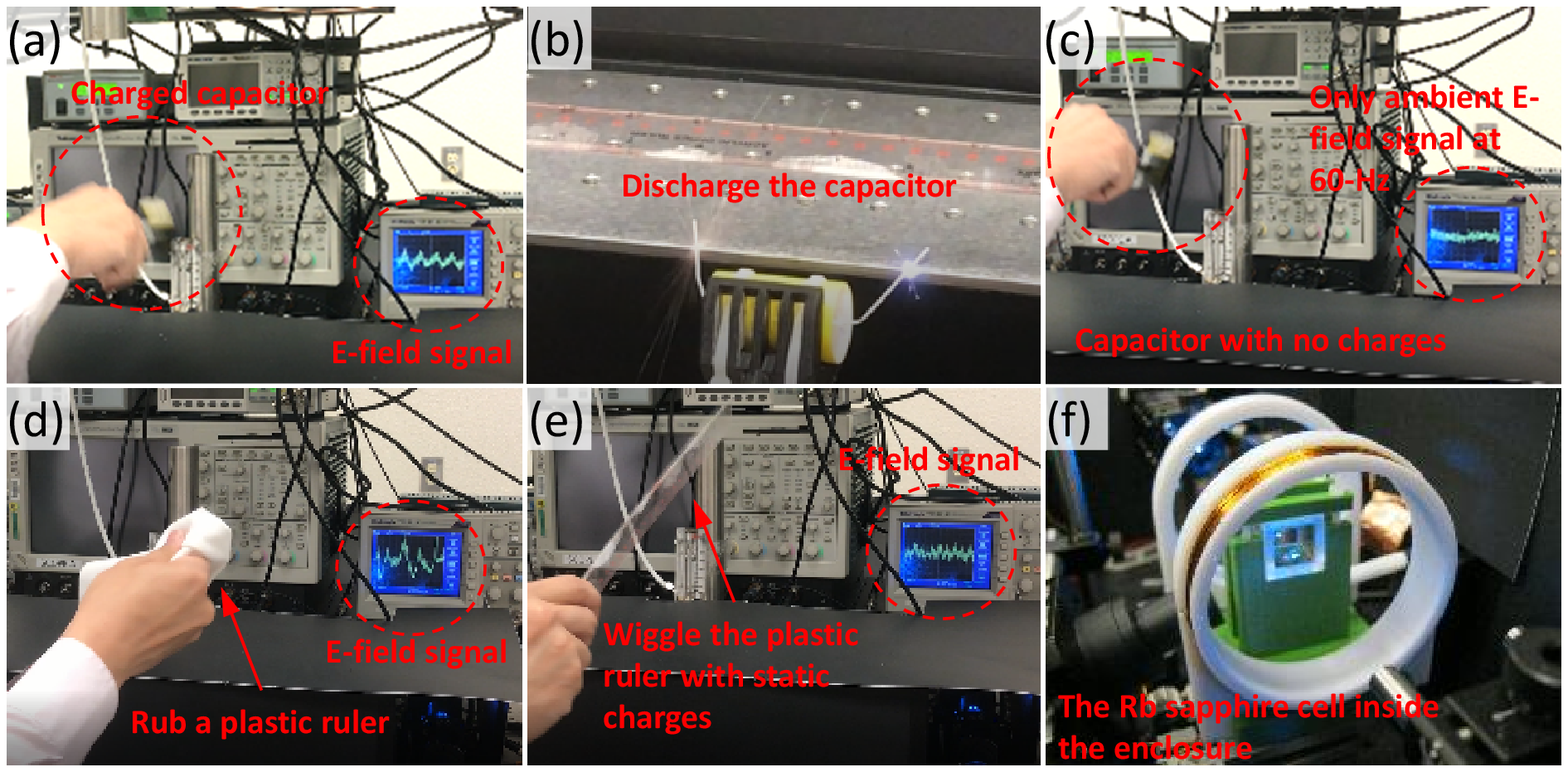}%
\caption{\label{VideoPhotos}Video snapshots: (a) Wiggle a charged capacitor above the vapor cell. (b) Discharge the capacitor. (c) No E-field signal was detected after discharging the capacitor. (d) Rub a plastic ruler. (e) Wiggle the plastic ruler with static charges. (f) Picture of the Rb sapphire cell mounted inside the enclosure.}
\end{figure*}
To characterize the spectral response, we set $I_{\rm LED}$ at 40 mA. We took time-dependent signal traces and performed Fourier analysis. The Fourier spectrum was then normalized using $|\tilde{\eta}(f)|$ defined by the measured $f_{\rm 3dB}$ to address the frequency-dependent E-field screening effect. In Fig.~\ref{Sensitivity}, we plot the measured E-field spectra with two different laser-power settings. For the case of $P_{780}=200$ $\mu$W and $P_{480}=10$ mW, the $f_{\rm 3dB}$ was determined to be $\approx64$ Hz. For the case of $P_{780}=200$ $\mu$W and $P_{480}=120$ mW, the $f_{\rm 3dB}$ was determined to be $\approx770$ Hz. We see these $f_{\rm 3dB}$ values are higher than the results deduced from the parameters listed in Table~\ref{CS}. This is due to the fact that the Rb vapor density and the 780-nm laser intensity in this experiment are higher than in the characterization experiment. With $\tilde{V}_{\rm noise}=0.65$ mV$/\sqrt{\rm Hz}$ at the output of the lock-in amplifier, which is about 1.2 times higher than the calculated photon shot noise, we find the E-field noise floor (at $f>f_{\rm 3dB}$, $\tilde{\eta}\approx1$) to be $|6.5\times10^{-3}/\beta\kappa|$. For $I_{\rm LED}=40$ mA and $P_{480}=120$ mW, we have $\kappa=-7.9$ MHz per V/m, and $\beta=0.24$ V/MHz, which gives a noise floor around 0.34 (mV/m)/$\sqrt{\rm Hz}$ as shown in Fig~\ref{Sensitivity}. The corresponding active volume for E-field sensing is about 11 mm$^3$ based on the beam diameter and 14 mm path length inside the vapor cell. We have verified that the influence of the laser frequency noise is insignificant to the detection noise floor for corresponding experimental conditions. Since $f_{\rm 3dB}\propto\gamma\propto P_{480}$, increasing $P_{480}$ can only improve the sensitivity at frequencies $\gtrsim f_{\rm 3dB}$ but not help sensing the E-field at frequencies $\ll f_{\rm 3dB}$, as we can see by comparing the spectrum of $P_{480}=10$ mW and the spectrum of $P_{480}=120$ mW in Fig.~\ref{Sensitivity}. Our experimental setup could only achieve 480-nm laser power at the cell up to 120 mW. If we are able to increase $P_{480}$ to 1 W, the E-field noise floor is expected to be $<0.1$ (mV/m)/$\sqrt{\rm Hz}$ with $f_{\rm 3dB}\approx6.4$ kHz, but there is only slight improvement regarding the sensitivity around 1-kHz detection frequency and basically no improvement for a detection frequency below 500 Hz as indicated by the black dashed curve in Fig.~\ref{Sensitivity}.

Using our low-frequency atomic electrometer, we showed that we can detect moving charged objects that are tens of centimeters away from the vapor cell. Figure~\ref{VideoPhotos} presents a few snapshots of a demonstration video showing the detection of a moving charged capacitor at 1000 V and a wiggled plastic ruler with static charges, and the associated E-field signals are displayed on an oscilloscope (see the video file in the supplementary material for more information). The ambient electric field at 60 Hz results in a faster oscillation on the oscilloscope signal, which leads to a fat looking trace on the oscilloscope in Fig.~\ref{VideoPhotos} due to the insufficient image resolution. To be noted, owing to the relatively high dielectric constant $\varepsilon\approx10$ of sapphire and the 3-mm thick cell walls, there is a dielectric shielding effect by about a factor of 2 based on the FEM calculation, and it is nearly frequency independent. Hence, the actual E-field amplitude and the sensitivity of the external E-field should be the measured signal amplitude and the noise floor multiplied by $\approx2$. But this dielectric shielding can be almost eliminated if we use low-dielectric glass materials with sapphire coating on the inner surface. For example, using 1-mm thick cell wall made of fused silica with sapphire coating, for the same cell dimension, the dielectric shielding factor will be $<1.1$.

\section{Summary and outlook}
We demonstrate low-frequency ($<$ kHz) atomic electrometry using a sapphire-made Rb vapor cell with internal bias E-field generated by the LED light. With the existing experimental setting, better E-field sensitivity at frequencies $>$ kHz can be achieved by simply increasing the 480-nm laser power or the vapor density. Although we are able to demonstrate detections of moving charged objects at a-few-Hz rate, the sensitivity level at a few Hz (see Fig.~\ref{Sensitivity}) is still much higher compared to the sensitivity level at a few hundred Hz. Therefore, the main challenge is how to further improve the E-field sensing performance at frequencies below hundreds of Hz. There are a few possible approaches for further improvement at lower frequencies. One is trying to produce a more uniform bias E-field $E_b$ and therefore the slope $\beta$ of the dispersive resonance signal will not decrease at higher $E_b$. This allows us to improve the overall sensitivity by continuing increasing the bias E-field until the $E_b$-dependent linewidth broadening due to other causes, such as energy-level mixing, take over. The other is to further reduce the E-field screening rate. In this work, we chose $100S_{1/2}$ state for performing atomic electrometry due to its very simple resonance structure that allowed us to obtain a decent dispersive-like resonance signal with non-zero bias E-field. Changing the EIT laser wavelengths to the combination of 795 nm and $\sim474$ nm for Rydberg-state interrogation can maximize the signal amplitude of a $nD_{3/2}$ resonance because of the higher oscillator strength and the minimized excitation to the $nD_{5/2}$ state due to the selection rule. A good dispersive-like $nD_{3/2}$ resonance may be obtained owing to its simpler structure compared to the $nD_{5/2}$ resonance. The same excitation probability for a transition from $5P_{3/2}$ to $nS_{1/2}$ can then be obtained using a transition from $5P_{1/2}$ to $n'D_{3/2}$, where $n'<n$. This implies that we may achieve the required $\kappa$ with lower blue laser power and therefore a lower E-field screening rate. We can also look for other materials for making a vapor cell or for a coating on the inner surface of the vapor cell that have even lower inner surface conductivity when exposing to alkali vapor. For example, some anti-spin-relaxation coating materials, such as paraffin, OTS (octadecyltrichlorosilane), etc., have shown a much lower adsorption energy, which may lead to a lower surface conductivity. Although these hydrocarbon coatings may have some adverse effects to the Rydberg states, they may still be worth studying. In summary, our work investigates a regime of vapor-cell-based atomic electrometry that was seldom explored before, and we have shown some encouraging results. Continuing this research direction will enable more applications using atomic E-field sensing technology.

\begin{acknowledgments}
We would like to thank Justin Christensen for his help in setting up some of experimental apparatus. We also thank David Meyer for his useful comments on the contents of this paper. This work was supported by the Laboratory Directed Research and Development (LDRD) program at Sandia National Laboratories (SNL). Sandia is a multimission laboratory managed and operated by National Technology and Engineering Solutions of Sandia (NTESS), LLC, a wholly owned subsidiary of Honeywell International Inc., for the U.S. Department of Energy National Nuclear Security Administration under contract DE-NA0003525. This paper describes objective technical results and analysis. Any subjective views or opinions that might be expressed in the paper do not necessarily represent the views of the U.S. Department of Energy or the United States Government.
\\
\end{acknowledgments}

\bibliography{LFAERefs}

\end{document}